\documentstyle[prb,multicol,aps]{revtex}
\input epsf

\begin{document}

\draft

\tighten

\preprint{MA/UC3M/19/1994}

\title{Nonlinear treatment of many-body effects in double-barrier\\
resonant tunneling structures}

\author{Enrique Diez and Angel S\'{a}nchez}

\address{Escuela Polit\'{e}cnica Superior, Universidad
Carlos III de Madrid, c./ Butarque 15, 
E-28911 Legan\'{e}s,
Madrid, Spain}

\author{Francisco Dom\'{\i}nguez-Adame}

\address{Departamento de F\'{\i}sica de Materiales,
Facultad de F\'{\i}sicas, Universidad Complutense, 
E-28040 Madrid, Spain}

\maketitle

\begin{abstract}

We introduce a simple, solvable model of double-barrier resonant
tunneling structure which includes the effects of electron-electron and
electron-phonon scattering.  The model is based on a generalized
effective-mass equation where a nonlinear coupling is introduced to
account for those inelastic scattering phenomena.  The nonlinear term
depends on one parameter which is the only constant in the model to be
determined phenomenologically from experimental data.  As an example of
the application of the model, we discuss GaAs-Ga$_{1-x}$Al$_x$As
double-barrier structures.  When the nonlinear term is considered, a
sideband is observed at an energy of the order of the LO phonon energy
in GaAs in addition to and below the main transmission peak.  Fitting of
the computed spectrum to known experimental results allows to find the
value of the nonlinear coupling.  As a consequence, we can estimate the
electron-phonon coupling from the shift of the main transmission peak in
comparison to the linear resonant tunneling.  We obtain good agreement
with values coming from more sophisticated treatments, which supports
our claim that the nonlinear term consistently includes electron-phonon
effects.  Finally, we use our simple model to study magnitudes of
interest for applications, namely $I-V$ characteristics, which are shown
to present two negative differential resistance peaks arising from the
main transmission peak and the sideband.

\end{abstract}


\begin{multicols}{2}

\narrowtext

\section{Introduction}

Resonant tunneling (RT) through double-barrier structures (DBS) make
these systems very promising candidates for a new generation of
ultra-high speed electronic devices.  For instance, a
GaAs-Ga$_{1-x}$Al$_x$As DBS operating at THz frequencies has been
reported in the literature.\cite{Sollner} The basic reason for RT to
arise in DBS is a quantum phenomenon whose fundamental characteristics
are by now well understood: There exists a dramatic increase of the
electron transmittivity whenever the energy of the incident electron is
close to one of the unoccupied quasi-bound-states inside the
well.\cite{Ricco} In practice, a bias voltage is applied to shift the
energy of this quasi-bound-state of nonzero width so that its center
matches the Fermi level.  Consequently, the $I-V$ characteristics
present negative differential resistance (NDR).

In actual samples, however, the situation is much more complex than this
simple picture.  This is so even in good-quality heterostructures, when
scattering by dislocations or surface roughness is negligible.  In
particular, {\em inelastic} scattering is always present in real
devices.  Examples of inelastic scattering events are electron-phonon
and electron-electron interactions, in which the energy of the tunneling
electron changes and the phase memory is lost.  The influence of these
many-body effects on DBS has recently attracted considerable attention.
Several approaches have been carried out to model
electron-optical-phonon scattering in RT devices.
\cite{Ned,Cai,Stovneng,Figielski} One thing that is currently clear is
that electron-phonon interaction shifts downwards the RT peak in energy
by $\Delta E=g\hbar\omega_0$, where $g$ is the electron-phonon coupling
and $\hbar\omega_0$ is the LO phonon energy.  It has also been noted
that sideband transmission peaks arise, which are readily observable in
$I-V$ characteristics, and correspond to sequential rather than coherent
tunneling.  On the other hand, the effects of Coulomb repulsion in RT
have also been considered in tight-binding models of
quasi-one-dimensional systems.\cite{Nonoyama} Results showed that the
conductance exhibits several doublet structures depending upon the $U$
coupling in the on-site Coulomb interaction between two electrons.
Also, each peak shifts upwards to higher energies as $U$ increases.
Therefore, we may summarize the available knowledge by saying that those
inelastic processes split the main RT peak and shift its energy
downwards or upwards, depending on the particular scattering channel.

Even with its rather satisfactory degree of success, many-body
calculations have difficulties that, in some cases, may complicate the
interpretation of the underlying physical processes.  For this reason,
an alternative approach to inelastic scattering effects above discussed
has recently emerged, based on nonlinear dynamics of single excitations
in solids.\cite{Davydov} Loosely speaking, this kind of treatment could
be regarded as similar to Hartree-Fock and other self-consistent
techniques, which substitute many-body interactions by a nonlinear
effective potential.  In this paper we present one of such a nonlinear
model, intended to account for many-body effects in RT through DBS
within the one-particle framework, by including nonlinear interaction
terms in the equation of motion.  We want to make it clear from the
beginning that we neither search for nor claim quantitative agreement
between our one-particle model and existing experiments.  We realize
that this is a shortcut also found in more elaborated many-body theories
and we cannot hope to cure this problem with a much simpler,
phenomenological theory: It is obvious that if we try to describe
electron-electron and electron-phonon interactions within the same
one-particle framework, quantitative accuracy is no longer possible.  On
the contrary, what we do claim is that our model captures the essential
physics of inelastic effects in RT in a very simple way.  The virtue of
such an approach is that it allows to gain insight on the features of
DBS without the burden of intensive computations providing, in an
inexpensive way, a qualitative picture of what is to be expected in
particular devices.  This is so because our model can be easily extended
to other systems in spite of the fact that we here present it in the
context of a specific DBS. Therefore, our model can give hints leading
to the development of sophisticated, possibly many-body calculations
with much better accuracy.  In addition, the way we connect effective
nonlinearity with the physical ingredients it tries to substitute may be
helpful for other nonlinear models in condensed matter physics.

The paper is organized as follows.  In Sec.~II, we present our model,
obtained by including a nonlinear coupling in a generalized
effective-mass equation.  We particularize it for a
GaAs-Ga$_{1-x}$Al$_x$As but we insist that the model is quite general
and applicable in different contexts.  The nonlinear coupling can be
regarded as an adjustable parameter to fit as well as possible the
available experimental results.  We discuss in detail the physics
underlying our choice for the coupling.  We sketch the exact solution of
this phenomenological model, as well as the way to obtain the
transmission coefficient as a function of
the nonlinear couplings and the applied voltage $V$.  For completeness,
we include a brief discussion of the range of applicability of the
equation and its connection with physical interpretations.  Afterwards,
Sec.~III contains the main results and discussions of our analysis
concerning electron transmission and $I-V$ characteristics.  We find the
correct value of the coefficient of the nonlinear coupling fitting
available data and, as a function of it, we discuss the corresponding
value of the electron-phonon coupling.  We compute the $I-V$
characteristics of the device and discuss its main features, the most
salient of them being the existence of two peaks of NDR. Finally,
Sec.~IV concludes the paper with a brief survey of the results, and some
prospects on the application of the ideas we have discussed.

\bigskip

\section{Model}

\subsection{Physical grounds and definitions}

To describe our model we have chosen a typical system: A
GaAs-Ga$_{1-x}$Al$_x$As DBS under an applied electric field.  The
thickness of the whole structure is $L$ and the thickness of the well is
$d$.  The barriers are assumed to be of the same thickness (symmetric
case) but as will be evident below this is not a restriction of our
approach.  The structure is embedded in a highly doped material acting
as contact, so that the electric field is applied only in the DBS. We
focus on electron states close to the bandgap and thus we can neglect
nonparabolicity effects hereafter.  Then the one-band effective-mass
framework is completely justified to get accurate results.  For the sake
of simplicity, we will further assume that the electron effective-mass
$m^*$ is constant through the whole structure.  This hypothesis is
related to the fact that we are not interested in high quantitative
accuracy, although we note that the spatial dependence
of the effective mass can
be taken into account if necessary.

Within this approach, the electron wave function is written as a sum of
products of band-edge orbitals with slowly varying envelope-functions.
Therefore the envelope-function $\psi(z)$ satisfies a generalized
effective-mass equation (we use units such that energies are measured in
effective Rydberg (Ry$^*$) and lengths in effective Bohr radius (a$^*$),
being $1\,$Ry$^*=5.5\,$meV and $1\,$a$^*=100\,$\AA\ in GaAs) given by

\begin{equation}
-\psi_{zz}(z)+\left[V(z)\chi_b(z)-eFz\right]\,\psi(z)=E\>\psi(z),
\label{Sch}
\end{equation}
where $V(z)$ is the potential term which we discuss below, $F$ is the
electric field, and $\chi_b(z)$ is the characteristic function of the
barriers,
\begin{equation}
\chi_b(z)=\left\{\begin{array}{ll} 1, & \mbox{\rm if}\ 0<z<(L-d)/2,\\
                                   1, & \mbox{\rm if}\ (L+d)/2<z<L,\\
                                   0, & \mbox{otherwise}.
                 \end{array} \right.
\label{chara}
\end{equation}

We now specify our model by choosing what is the potential term $V(z)$.
In order to do that, let us first consider the physics we are trying to
represent with this term.  The DBS can be regarded as an effective
medium which reacts to the presence of the tunneling electron, leading
to a feedback mechanism by which inelastic scattering processes change
the RT characteristics of the device.  It thus follows that $V(z)$ must
contain nonlinear terms if it is to summarize the medium reaction which
comes from the electron-electron and electron-phonon interaction.  The
simplest candidate to contain this feedback process is the charge
density of the electron, which is proportional to $|\psi(z)|^2$.  In our
model, we neglect higher order contributions and postulate that the
potential in Eq.~(\ref{Sch}) has the form
\begin{equation}
\label{Veff}
V(z) \equiv V_b\left[1+\tilde{\alpha}|\psi(z)|^2 \right],
\end{equation}
where $V_b$ is the conduction band-offset at the interfaces, and all the
nonlinear physics is contained in the coefficient $\tilde{\alpha}$ which
we discuss below.

There are two factors
that configure the medium response to the tunneling electron.  First, it
goes without saying that there are repulsive electron-electron Coulomb
interactions, which should enter the effective potential with a positive
term proportional to the charge, i.e., the energy is increased by local
charge accumulations, leading to a positive sign for $\tilde{\alpha}$.
On the other hand, in polar semiconductors, the electron polarizes the
surrounding medium creating a local, positive charge density.  Hence the
electron reacts to this polarization and experiences an attractive
potential (which implies $\tilde{\alpha}$ negative).
This happens, for instance, in the polaron problem in the
weak coupling limit, which becomes valid in most semiconductors, and
where it can be seen that the lowest band energy state
decreases.\cite{Callaway} 
It is then clear that in principle any sign
would be equally possible for this coefficient if $\tilde{\alpha}$ is to
represent the combined action of the polarization of the lattice along
with repulsive electron-electron interactions.
Intuitively, however, it is most realistic to think
that $\tilde{\alpha}$ will be
negative, because a positive nonlinear interaction would arise from
negative charge accumulation in the barriers, which is not likely to
occur.  In fact, we discuss below mathematical reasons imposing that
$\tilde{\alpha}$ has to be
negative as expected, allowing for it to be positive only if it is 
small and the barriers themselves are very narrow.
It can be argued following this line of reasoning
that charge accumulation is expected inside the well, and that we should
include a nonlinear term in the well to account for that.  We have
studied this possibility as well and we found that such a term is
irrelevant, as we explain in detail in Sec.\ III.

\subsection{Analytical results}

In the preceding subsection, we have defined our model, and the only
remaining thing is to fix the value of $\tilde{\alpha}$.  This we
address in the next section.  We now work starting from Eq.\ (\ref{Sch})
with the definition in Eq.\ (\ref{Veff}) to cast the equations in a more
tractable form.  For simplicity, and because we are interested in
intrinsic DBS features, we consider that the contacts in which the
structure is embedded behave linearly.  Therefore, the solution of
Eq.~(\ref{Sch}) is a linear combination of traveling waves.  As usual in
scattering problems, we assume an electron incident from the left and
define the reflection $r$ and transmission $t$ amplitudes by the
relationships
\begin{equation}
\psi(z)=\left\{ \begin{array}{ll} A\left(e^{ik_0z}+re^{-ik_0z}\right)
       & z<0, \\ Ate^{ik_Lz} & z>L,  \end{array} \right.
\label{solution}
\end{equation}
where $k_0^2=E$, $k_L^2=E+eFL$, and $A$ is the incident wave amplitude.
Now we define $\psi(z)=A\phi(z)$, and $\alpha=\tilde{\alpha}|A|^2$.
Notice that $\alpha$ is a dimensionless parameter.  Using
Eq.~(\ref{Sch}) we get
\begin{equation}
-\phi_{zz}(z)+\left\{V_b\chi_b(z)\left[1+\alpha
|\phi(z)|^2\right]-eFz-E\right\}\,\phi(z)=0,
\label{fi}
\end{equation}

To solve the scattering problem in the resonant structure we develop a
similar approach to that given in Ref.~\onlinecite{Knapp}.  Since
$\phi(z)$ is a complex function, we take $\phi(z)=q(z)\exp[i\gamma(z)]$,
where $q(z)$ and $\gamma(z)$ are real functions.  Inserting this
factorization in Eq.~(\ref{fi}) we have $\gamma_z(z)=q^{-2}(z)$ and
\begin{eqnarray}
-q_{zz}(z)+{1\over q^3(z)}+\left[V_b\chi_b(z)-eFz-E\right]\,q(z) & + &
\nonumber \\
+\alpha V_b\chi_b(z)q^3(z) & = & 0.
\label{q}
\end{eqnarray}
This nonlinear differential equation must be supplemented by appropriate
boundary conditions. However, using Eq.~(\ref{solution}) this problem
can be converted into a initial conditions equation. In fact, it is
straightforward to prove that 
\begin{equation}
\label{ic}
q(L)=k_L^{-1/2},\>q_z(L)=0,
\end{equation}
and that the transmission coefficient is given by
\begin{equation}
\tau=\,{4k_0q^2(0)\over 1+2k_0q^2(0)+k_0^2
q^4(0)+q^2(0)q_z^2(0)}.
\label{tau}
\end{equation}
Hence, we can integrate numerically (\ref{q}) 
with initial conditions (\ref{ic}) backwards,
from $z=L$ up to $z=0$, to
obtain $q(0)$ and $q_z(0)$, thus computing the transmission coefficient
for given nonlinear coupling $\alpha$, incoming energy $E$
and applied voltage $V=FL$.

Once the transmission coefficient has been computed, and recalling that
contacts are linear media, the tunneling current density at a given
temperature $T$ for the DBS can be calculated within the
stationary-state model from
\begin{mathletters}
\label{eq2}
\begin{equation}
j(V)={m^*ek_BT\over 2\pi^2\hbar^3}\,\int_0^\infty\> \tau(E,V)N(E,V)\,dE,
\label{eq2a}
\end{equation}
where $N(E,V)$ accounts for the occupation of states to both sides of
the device, according to the Fermi distribution function, and it is
given by
\begin{equation}
N(E,V)=\ln\left(\frac{1+\exp[(E_F-E)/k_BT]}{1+\exp[(E_F-E-eV)/k_BT]}
\right),
\label{eq2b}
\end{equation}
\end{mathletters}
where $k_B$ is the Boltzmann constant.

\subsection{Sign of the nonlinear term}

At this point, we are in a position to give mathematical reasons why
$\alpha$ (or $\tilde{\alpha}$) should be negative by using
Eq.~(\ref{q}).  This equation has the form
\begin{equation}
\label{q2}
-q_{zz}(z)+{1\over q^3(z)}+f_1(z)q(z)+f_2(z)q^3(z)=0,
\end{equation}
where $f_1(z)$ and $f_2(z)$ are well-behaved functions.  Let us now
consider the discrete version of this equation, with the second
derivative discretized in the usual way; denoting $q_n=q(z=n\Delta z)$
and $f_{in}=f_i(n\Delta z)$, $i=1,2$, with $\Delta z$ being the integration
step, Eq.~(\ref{q2}) can be rewritten as
\begin{equation}
\label{q3}
q_{n+1} = 2q_n - q_{n-1}+(\Delta z)^2 \left[ {1\over q_n^3} + f_{1n} q_n
+ f_{2n}q_n^3\right].
\end{equation}

Notice that in this expression the sign of $f_{2n}$ is the same as the
sign of $\alpha$ and $\tilde{\alpha}$.  Let us now consider
Eq.~(\ref{q3}) for large $q_n$; this is general, because if $q$ is small
the term $q_n^{-3}$ will make it grow quite quickly.  In this limit,
Eq.~(\ref{q3}) can be approximately replaced by
\begin{equation}
\label{q5}
\Delta q_{n-1} = \Delta q_n + (\Delta z)^2 f_{2n}q_n^3,
\end{equation}
where $\Delta q_n=q_n-q_{n+1}$, and we have cast the equation in this fashion
because it is to be integrated backwards. Recalling the initial conditions 
(\ref{ic}), if $N$ is the total number of grid points, we have $q_N=
q_{N-1}=k_L^{-1/2}>0$. Therefore, 
we see that $\Delta q_{N-1}=0$, and 
\begin{equation}
\label{eh}
\Delta q_{N-2}=(\Delta z)^2 f_{2n}
k_L^{-3/2}.
\end{equation}
We thus see that if $f_{2n}$ is positive
(and hence $\alpha$) the first increment is positive, and so are the 
subsequent ones,
leading to a exponential divergence of $q$,
whereas if $f_{2n}$ and $\alpha$ are negative, the increment is
negative, and $q$ decreases until the $q_n^{-3}$ starts being relevant
again.  In this last case, it is
possible that $q$ reaches an equilibrium due to the
balance of the two cubic terms, which was not for $\alpha>0$.
This is in fact seen in the numerical integration of
Eq.~(\ref{q}), where $q$ rapidly diverges if $\alpha$ is positive,
unless, of course, $\alpha$ is positive but very small and/or the 
barrier (the region for $\alpha$ to influence $q$) is very narrow. In 
this last situation, however, the effect of $\alpha$ becomes negligible.

Physically, this can be understood as follows.  Electrons impinge on the
barrier from outside, and begin to tunnel through it, their wavefunction
being real and exponentially increasing or decreasing.  If $\alpha$ is
positive, then if there were any charge density in the BDS, this would
become even more repulsive, and the wavefunction would diverge even
faster (numerically one always see the exponentially growing part, of
course).  This instability is not present in the opposite case, where a
negative $\alpha$ helps the electron tunnel across the barrier. 

\section{Results and discussions}

In our calculations we have considered a double-barrier
GaAs-Ga$_{0.65}$Al$_{0.35}$As structure with $L=3d=150\,$\AA.  The
conduction-band offset is $V_b=250\,$meV.  In the absence of applied
electric field and nonlinearities, there exist a single, very narrow
resonance $\tau\sim 1$ below the top of the barrier, with an energy of
$80.7\,$meV.  
\begin{figure}
\setlength{\epsfxsize}{5.2cm}
\centerline{\mbox{\epsffile{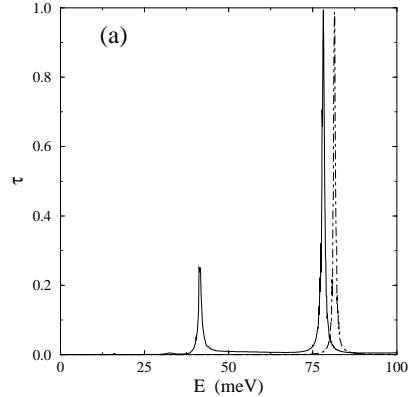}}}
\setlength{\epsfxsize}{5.2cm}
\centerline{\mbox{\epsffile{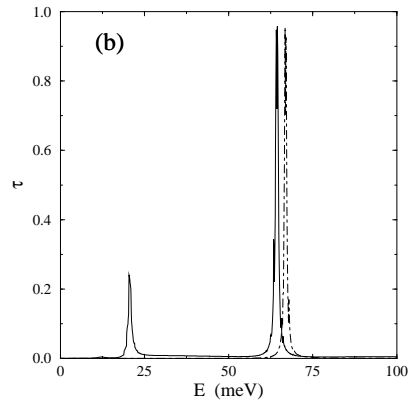}}}
\setlength{\epsfxsize}{5.2cm}
\centerline{\mbox{\epsffile{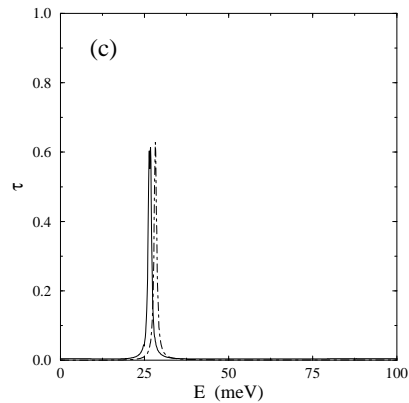}}}
\caption{Transmission coefficient $\tau$ as a function of the
electron energy for $\alpha=-7.6\times 10^{-3}$,
for different values of the applied voltage (a) $0\,$volts,
(b) $0.029\,$volts, and (c) $0.105\,$volts. For comparison, dashed lines
indicate the results for $\alpha=0$.}
\label{fig1}
\end{figure}

Hence the well supports a single quasi-bound state.  When
there exists an applied voltage, the energy of the quasi-bound state
level is lowered and a strong enhancement of the current arises whenever
the Fermi level matches this resonance, thus leading to the well-known
RT phenomenon.  This picture changes dramatically if a small amount of
nonlinearity appears in the structure.

Figure~\ref{fig1} shows the transmission coefficient as a function of
the incoming energy for different values of the applied bias, when
$\alpha=-7.6\times 10^{-3}$.  This value is the one that best fits the
experimental splitting that we discuss now.  At zero bias, shown in
Fig.~\ref{fig1}(a), there exist two well differentiated peaks, the main
one centered at $78.2\,$meV and the sideband centered at $41.5\,$meV.
The main peak not only shifts to lower energies but also broadens, in a
similar fashion to what happens in RT when inelastic effects arise.  The
energy separation between the main peak and the sideband is $36.7\,$meV,
very close the energy of LO phonons in GaAs ($\hbar\omega_0 =
36.3\,$meV), which was our requirement to choose $\alpha$.  Therefore,
the self-interaction we introduce in our model gives results similar to
those found in dealing with the full electron-phonon Hamiltonian in DBS,
\cite{Cai} where the sideband is due to phonon effects at inelastic
channels.  Interestingly, at the sideband the transmission probability
is close to $0.2$, which agrees with more sophisticated
treatments.\cite{Cai} This agreement reinforces our hope to be treating
in a proper way these inelastic scattering effects.  Furthermore, to
check the validity of our model, we can estimate the electron-phonon
coupling from the downwards shift of the main peak when nonlinearities
are introduced.  This shift is $\Delta E = 2.8\,$meV, and using the
value of $\hbar\omega_0=36.7\,$meV obtained previously we get $g=0.076$,
in rather good agreement with the value $g=0.10$ proposed by Cai {\em et
al.\/} \cite{Cai} We note, however, that this result is not far from
$g=0.03$ given by Wingreen {\em et al\/.}\cite{Ned} In fact, both
many-body results enclose ours, and our calculation is not clearly
favorable to any of them.

We now consider the effect of a bias imposed on the DBS. When there exists
an applied voltage, the transmission peaks also shifts to lower energies
due to the presence of the linear potential, as shown in
Fig.~\ref{fig1}(b).  This behavior is similar to that found in linear
(coherent) RT. For instance, these peaks are located at $62.9\,$meV and
$15.2\,$meV for $V=0.029\,$volts.  A further increase of the applied
voltage makes the sideband to disappear [see Fig.~\ref{fig1}(c)].  The
reasons for that will be explained below.

\begin{figure}
\vspace*{0.3 in}
\setlength{\epsfxsize}{8cm}
\centerline{\mbox{\epsffile{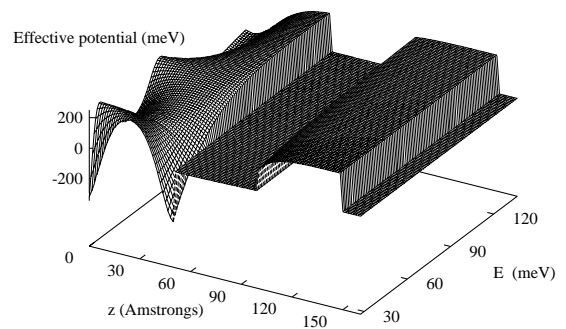}}}
\caption{Effective potential $V_{ef\!f}$ as a function of $z$ and the
incoming electron energy $E$ at zero bias, for $\alpha=-7.6\times
10^{-3}$.}
\label{fig2}
\end{figure}

In order to gain insight on the nonlinear RT in DBS, we can rewrite
Eq.~(\ref{fi}) as follows
\begin{mathletters}
\label{defi}
\begin{equation}
-\phi_{zz}(z)+V_{ef\!f}(z,E)\,\phi(z)=E\>\phi(z),
\label{defia}
\end{equation}
where we have defined an effective potential as follows
\begin{equation}
V_{ef\!f}(z,E)=V_b\chi_b(z)[1+\alpha q^2(z)]-eFz.
\label{defib}
\end{equation}
\end{mathletters}
Thus (\ref{defi}) is a Schr\"odinger-like equation for an effective
potential due to nonlinearity plus the linear potential and the built-in
potential of the DBS. This effective potential depends not only on $z$
but also on the incoming energy $E$ through the function $q(z)$.  Since
the envelope function changes under RT conditions, and also $q(z)$
accordingly, it should be expected that $V_{ef\!f}(z,E)$ undergoes
severe variations whenever $E$ is close to one of the RT peaks.  This is
indeed the case, as shown in Fig.~\ref{fig2}, where $V_{ef\!f}(z,E)$ is
displayed at zero bias.  Notice that nonlinear effects have negligible
effects on the shape of the effective potential in the right barrier,
besides a slight band bending at the interface $z=(L+d)/2$ at low
energies.  However, the potential in the left barrier region differs
significantly from the original square-barrier shape.  Out of
resonances, the effective
barrier height at $z=0$ is lower than $V_b$, whereas at
resonances it takes the value $\sim V_b$.  Hence the effective potential
presents two local maxima in the plane $z=0$ as a function of the
incoming energy $E$, matching the values of the main resonance and the
sideband above discussed.  From Fig.~\ref{fig2} it is clear that at the
energy of the main resonance the effective potential is quite similar to
the built-in DBS potential, just producing a small shift of the
quasi-bound-state in comparison to the linear RT process.  Concerning
the sideband, the effective potential presents a deep minimum at the
interface $z=(L-d)/2$, which causes the lowering of the
quasi-bound-state, thus explaining the origin of this lower RT peak.
Indeed, the fact that this added well is responsible for the peak is
confirmed by the observation that the disappearance of the peak as the
field increases coincides with the vanishing of the well due to the
bias; before that, the resonance responsible for the peak is moving
towards $E=0$ where it can be seen that the depth of the 
secondary well vanishes.  We thus provide a
complete, coherent picture of the tunneling phenomenology.

\begin{figure}
\setlength{\epsfxsize}{5.8cm}
\centerline{\mbox{\epsffile{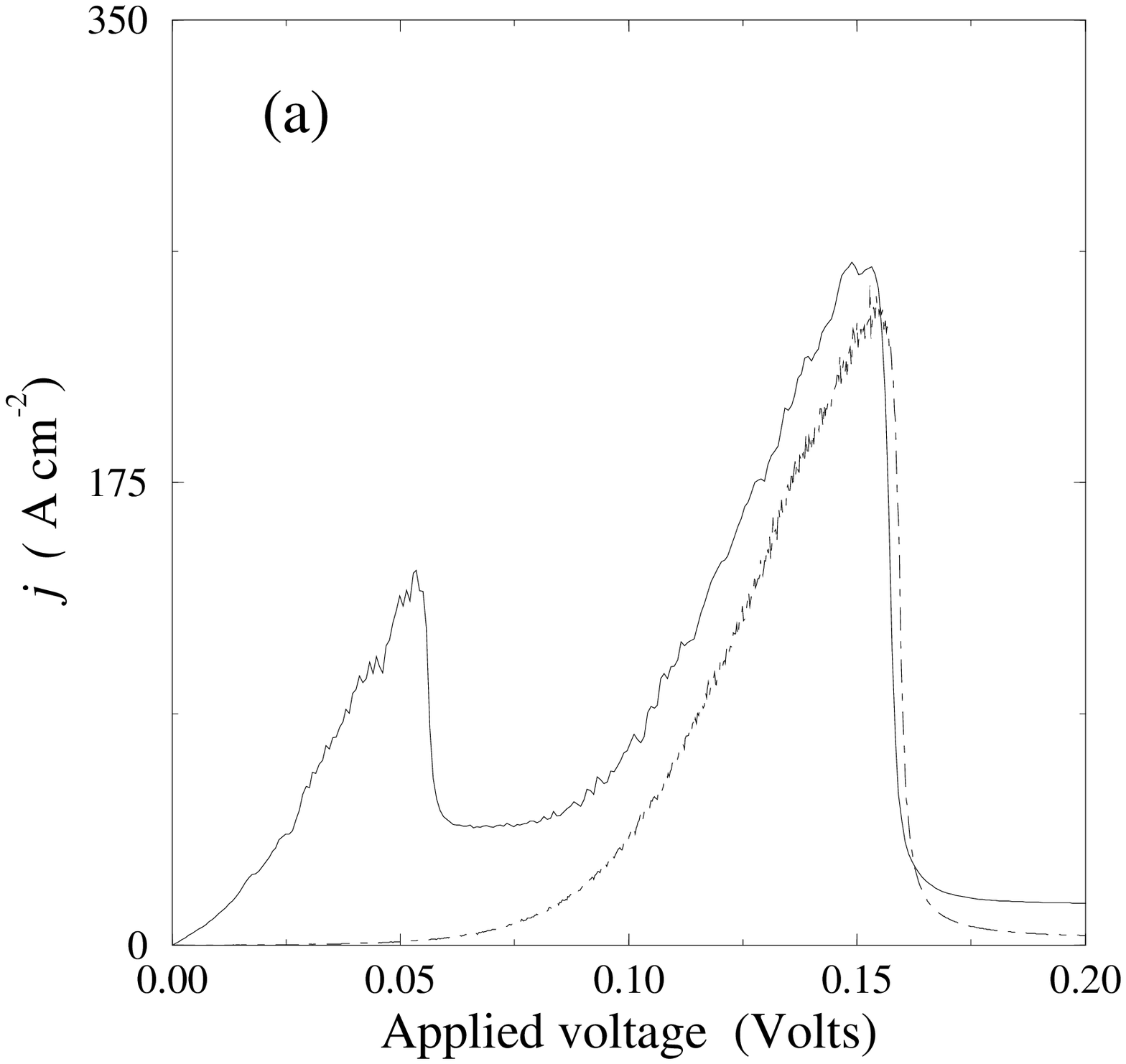}}}
\setlength{\epsfxsize}{5.8cm}
\centerline{\mbox{\epsffile{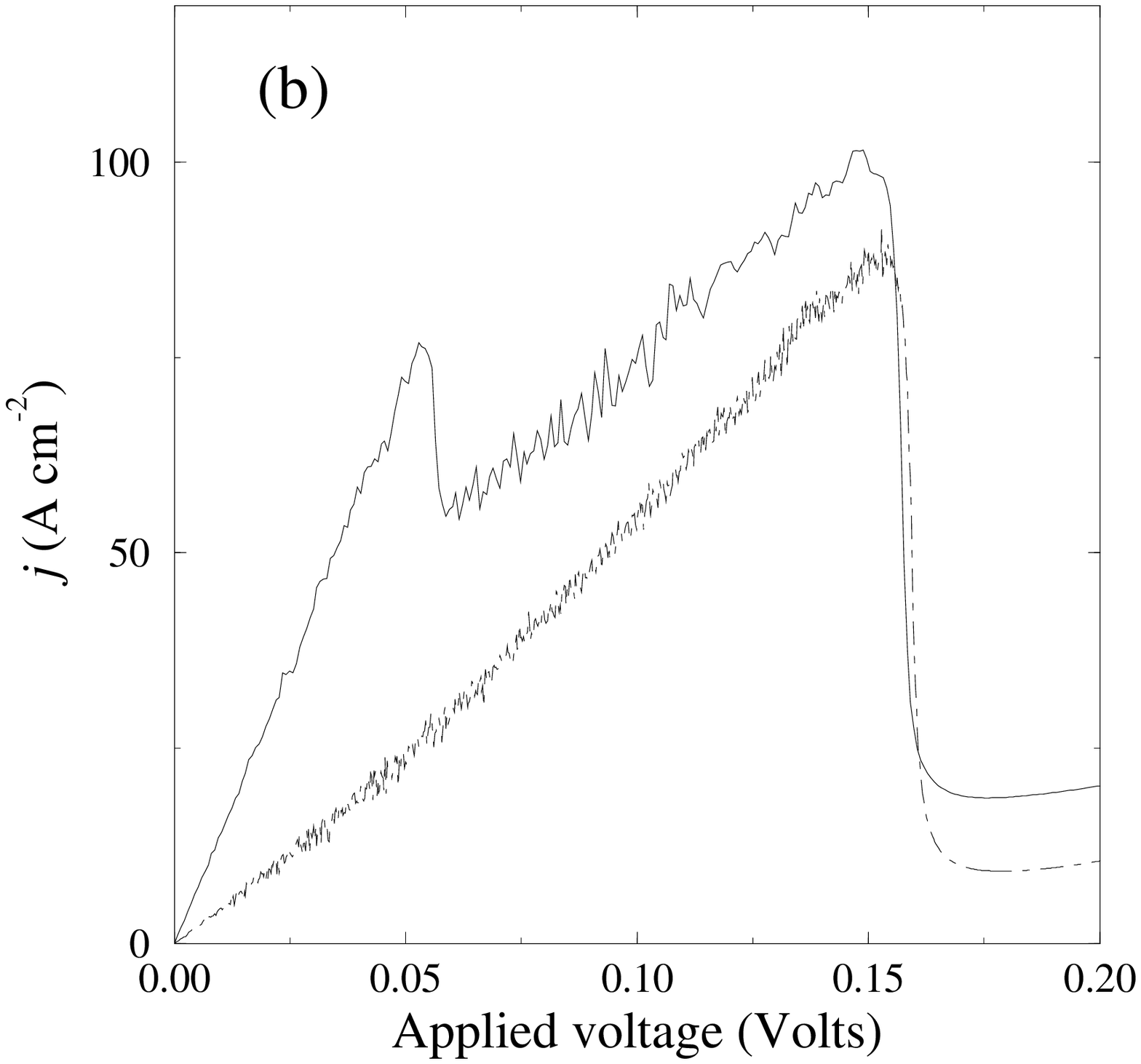}}}
\caption{Computed $j-V$ characteristics for $E_F=55\,$meV,
$\alpha=-7.6\times 10^{-3}$  at (a)
$T=77\,$K and (b) room temperature. For comparison, dashed lines
indicate the results for $\alpha=0$.}
\label{fig3}
\end{figure}

We now have to comment the $j-V$ characteristics, computed from
(\ref{eq2}).  We have set two different temperatures ($77\,$K and room
temperature) and compared these curves with those obtained in linear RT.
The Fermi energy was $E_F=27.5\,$meV.  Results are shown in
Fig.~\ref{fig3}.  In all cases we obtained clear NDR signatures.  The
linear RT shows a single NDR peak whereas self-interaction causes the
occurrence of a second peak at a lower voltage, clearly related to the
sideband in the transmission coefficient.  On increasing temperature
both peaks merges into a single one because of the broadening of the
Fermi-Dirac distribution function.  It is important to notice that the
$j-V$ characteristics collect all the main features found in the
transmission coefficient, namely shift downwards of the main RT peak,
its corresponding broadening due to inelastic effects and the appearance
of the sideband, shown in Fig.~\ref{fig3} as a smaller NDR signature at
$V=0.05\,$volts. 

Finally, some words are in order about possible nonlinearities in the
well, as we announced in Sec.\ II. In principle, charge may accumulate
in the well depending upon doping degree, and that is expected to induce
a self-repulsive interaction of the electronic cloud, overcoming the
attractive part due to polarization of the medium.  So, if we are
considering that nonlinearities are zero in the well we could argue that
one effect compensates the other and then there is no nonlinear term to
be added.  Admittedly, that would be a very rare phenomenon, because
exact compensation of both forces (or any interaction in nature) would
be serendipitous.  The question then arises as to whether our model is a
good one in the sense that it is robust against new terms accounting for
inelastic scattering in the well.  We checked that by introducing
another term in the potential, given by
$V'(z)\equiv\tilde{\beta}|\psi|^2$ for $(L-2)/2<z<(L+d)/2$.  The
calculations can be carried out much in the same way we have discussed,
and the results obtained by numerical integration of an equation very
close to Eq.\ (\ref{q}).  We studied a range of values for
$\beta\equiv\tilde{\beta}|A|^2$ of the same order as those of $\alpha$:
$\beta$ positive would
indicate that Coulomb effects dominate, whereas $\beta$ negative would
be the same as in the barriers, a representation of the effects of
polarization.  The results were very much satisfactory, because the
features of the model did not change, and only minor quantitative
modifications were found.  We can then conclude that leaving the
possible nonlinearity of the well out of our model is not a limitation
and that it is not necessary to take it into account.

\section{Conclusions}

In conclusion, we have presented a nonlinear effective-mass equation to
take into account in a simple way electron-electron and electron-phonon
interaction in RT through DBS. Our model represents those effects by
including a self-coupling of the wavefunction with the charge cloud
inside the barriers, whose strength is the only adjustable parameter.
By comparing the results of the model to experimental data on the
distance between the main transmission peak and the sideband we fix the
value of that constant. 
With the same value we found transmission
probabilities and electron-phonon coupling values very close to those
arising from many-body calculations, which indicates that our model
consistently reproduces their features. We have also been able to
discard any possible nonlinearity in the well, which shows that our
understanding of the physics of this problem is in the right direction.
We have to stress that the low values obtained for the nonlinear
coupling, $\alpha=-7.6\times 10^{-3}$, do agree with our purpose of
offering an alternative description: Had we obtained large values of
$\alpha$, it would be very difficult to understand how those magnitudes
could arise from what in principle are perturbative effects of the
linear description.  In addition, large values of $\alpha$ would have
immediately posed the question about the role that soliton formation and
transmission could be playing along with RT in the DBS (see
Refs.~\onlinecite{Knapp} and \onlinecite{yomismo} as well as references
therein).  Therefore, our model satisfactorily achieves the goals we had
in mind, a great success if its simplicity is taken into account.

Further extensions of the present work to study nonlinear dynamical
response of DBS on external ac bias would be of great interest to shed
light on related problems like bistability, \cite{Pawel} noise
characteristics, \cite{Flores} and RT at far infra-red frequencies
\cite{Chitta} under the influence of inelastic scattering channels as
those described here.  Besides that, the above mentioned highly
nonlinear limit could also be interesting, for nonlinearity is always
susceptible to give rise to new and unexpected features.  If these new
features were seen in our model it would be a very exciting development.
If materials with suitable characteristics were found, and that is to be
expected, experiments could be made to check the predictions: If the
model were wrong, that would establish its range of validity, whereas if
the predictions were correct, this work could pave the way to a new
family of devices and applications.

\acknowledgments

We are very thankful to Paco Padilla for useful discussions.
A.\ S.\ acknowledges partial support from C.I.C.\ y T.\ (Spain) through
project No PB92-0248 and by the European Union Human Capital and
Mobility Programme through contract ERBCHRXCT930413.  F.\ D-A.\
acknowledges support from UCM through project No PR161/93-4811.

\end{multicols}

\end{document}